\documentclass{pasj00}

\draft

\begin{document}
\SetRunningHead{S. Kato}{}
\Received{2011/00/00}%{yyyy/mm/dd}
\Accepted{2011/00/00}%{yyyy/mm/dd}

\title{Correlated Frequency-Changes of Trapped Vertical p-mode Oscillations and kHz QPOs}

%%% begin:list of authors
\author{Shoji \textsc{Kato}}
    %\thanks{Example: Present Address is xxxxxxxxxx}}
\affil{2-2-2 Shikanodai-Nishi, Ikoma-shi, Nara, 630-0114}
\email{kato.shoji@gmail.com, kato@kusastro.kyoto-u.ac.jp}
%\and
%\author{C-Firstname {\sc C-Familyname}}
%\affil{C-Address of Institute}\email{ccccc@xxx.xxx.xx.xx}
%%% end:list of authors
%%% Please use the following style in case that sorting by 
%%% affilation is impossible. 
%
% \author{%
%   D-Firstname \textsc{D-Familyname}\altaffilmark{1}
%   E-Firstname \textsc{E-Familyname}\altaffilmark{1,2}
% and
%   F-Firstname \textsc{F-Familyname}\altaffilmark{2}}
% \altaffiltext{1}{Address of Institute}
% \email{ddddd@xxx.xxx.xx.xx}
% \email{eeeee@xxx.xxx.xx.xx}
% \altaffiltext{2}{Address of Institute}

%% `\KeyWords{}' always has to be placed before `\maketitle'.
 %Do NOT move this preamble from here!

\maketitle

\begin{abstract}
We have examined the frequencies of trapped two-armed ($m=2$) nearly vertical
oscillations (vertical p-mode) in vertically isothermal disks with toroidal
magnetic fields.
The magnetic fields are stratified so that the Alfv\'{e}n speed, $c_{\rm A}$,
is constant in the vertical direction.
We have particularly focused our attention on how frequencies of the fundamental
mode ($n_{\rm r}=0$) and first overtone ($n_{\rm r}=1$) in the radial direction change with 
correlation, when the ratio $c_{\rm A}^2/c_{\rm s}^2$ changes, $c_{\rm s}$ being the isothermal 
acoustic speed.
The results show that in the case where the oscillations are fundamental mode
($n=1$) in the vertical direction, the correlated frequency changes of the above-mentioned
oscillations seem to well describe, with standard values of
the mass and spin of the central sources, the frequency correlation of kHz QPOs
observed in neutron-star X-ray binaries.   
\end{abstract}

\section{Introduction}

Many authors now think that the kilo-hertz quasi-periodic oscillations (kHz QPOs) observed in low-mass
X-ray binaries (LMXBs) are some kinds of disk oscillations in a strong gravitational field.
Understanding their origin is of importance, since it will become a promising tool to evaluate the mass and spin of
the central sources as well as the disk structure surrounding the sources.  
Many models of kHz QPOs related to disk oscillations and waves have been proposed.
None of them, however, can sufficiently well describe the observed frequency correlation of the twin kHz QPOs
(Lin et al. 2010).

As one of possible models of kHz QPOs, for example, we proposed a disk-oscillation model in
deformed disks (Kato and Fukue 2006), based on the fact that in deformed disks
a set of oscillations with positive and negative wave energies are excited resonantly
(Kato 2004, 2008a,b; Ferreira and Ogilvie 2008; Oktariani et al. 2010; Kato et al. 2011; Kato 2011b).
This model qualitatively describe the frequency correlation of observed twin QPOs.
However, there are some problems.
The major problems are that the mass required to describe the observed frequency correlation of the twin
QPOs of neutron-star X-ray binaries is as large as $2.4M_\odot$ and further that the calculated correlation 
deviates gradually from the observed
one as frequency increases (Kato 2007; Lin et al. 2010).

Other disk oscillation models we have are those in which trapped oscillations are directly connected to
QPOs.
In geometrically thin disks, there are many kinds of disk oscillation modes.
In terms of the node number in the vertical direction, $n$, and the 
frequency in the corotating frame, these oscillation modes are classified into four, i.e.,
p-, g-, c-, and vertical p-modes (for details of the classification, see Kato 2001, Kato et al. 2008).
Some oscillations of these modes are known to be trapped in the innermost region of relativistic disks,
and thus they become a set of oscillations with discrete frequencies.
These trapped oscillations have been studied by many authors, e.g.,
by Kato and Fukue (1980), Ortega-Rodr\'{i}gues et al. (2002), Lai and Tsang (2009) for p-mode oscillations;
by Okazaki et al. (1987), Nowak and Wagoner (1992), Perez et al. (1997), Fu and Lai (2009) for g-mode oscillastions; by 
Kato (1990), Silbergleit et al. (2001) for c-mode oscillations; and by Kato (2010, 2011a) 
for the vertical p-mode oscillations (see Wagoner 1999, Kato 2001, and 
Kato et al. 2008 for reviews of trapped oscillations). 
There seems, however, to be no quantitative studies examining whether a set of these oscillations can really 
describe the correlated frequency-changes  of twin kHz QPOs.
Here, we discuss this problem, by especially focusing on the vertical p-mode oscillations, using the results obtained by
Kato (2011a).
%Finally, an application of the vertical p-mode oscillations to high frequency QPOs of black-hole X-ray binaries  
%is briefly  mentioned.

\section{Trapped Vertical p-mode Oscillations in Magnetized Disks}

Here, we focus our attention on oscillations which have at least one node, $n\geq 1$, 
in the vertical direction.\footnote{
When we mention the node number in the vertical direction, it is the number of node(s) of the
density perturbation in the vertical direction.
In the case of $n=1$, the density perturbation has a node on the equatorial plane.
It is noted that the node number associated with the vertical component of velocity is smaller than
that of the density perturbation by one.
}
There are two oscillation modes with a given $n$, when $n\geq 1$.
One is the g-mode, the square of whose frequencies in a corotating frame, $(\omega-m\Omega)^2$,
is smaller than the square of the epicyclic frequency, $\kappa^2$,
i.e., $(\omega-m\Omega)^2<\kappa^2$, where $\omega$ is the frequency of oscillations in the inertial frame,
$m$ is the azimuthal wavenumber and $\Omega$ is the angular velocity of the disk rotation.
The other mode of oscillations with $n\geq 1$ is the vertical p-mode, the square of whose frequencies in the
corotating frame is high in the sense that $(\omega-m\Omega)^2>\Omega_\bot^2$, where $\Omega_\bot$
is the vertical epicyclic frequency and always larger than $\kappa$.
In this paper we consider the vertical p-mode oscillations of $n=1$ and $2$ with $m=2$.

To understand characteristics of trapping of the vertical p-mode oscillations, let us first consider the case where
the disk is vertically isothermal and has no global magnetic fields.
Then, the propagation region of the vertical p-mode oscillations with $\omega$, $m$, and $n$ is specified by
$(\omega-m\Omega)^2>n\Omega_\bot^2$ (e.g., see Kato 2001).
This means that the propagation regions are divided into two radial regions specified by 
$\omega<m\Omega-n^{1/2}\Omega_\bot$ and $\omega>m\Omega+n^{1/2}\Omega_\bot$.\footnote{
Among oscillations of $(\omega-m\Omega)^2>\Omega_\bot^2$, one-armed ($m=1$) one with one node ($n=1$) in the vertical
direction has a particular position.
The oscillation is called c-mode (corrugation mode).
It is nearly incompressible motions changing disk plane up and down with a corrugation pattern.
This mode is outside of our interest in this paper.
}
The former shows that the propagation region of prograde oscillations, $\omega>0$, with $m=2$ and $n\leq 4$ is limited only in 
the innermost region of disks, and these oscillations are trapped there.
That is, we can expect a discrete set of trapped oscillations.

Now we consider the case where global toroidal magnetic fields exist in disks.
Since the vertical p-mode oscillations are oscillations of nearly vertical motion, they are the fast mode of 
three MHD waves, when toroidal magnetic fields exist.
Hence, the square of the frequency in the corotating frame becomes larger than $n\Omega_\bot^2$ by the
effects of magnetic restoring force, say $(\omega-m\Omega_\bot)^2>(n+\alpha)\Omega_\bot^2$, where
$\alpha$ is a dimensionless quantity of the order of unity, depending on $n$ and the strength of magnetic fields
(see Kato 2011a for details).
Hence, the propagation region near the inner region of the disks is changed to
$\omega<m\Omega-(n+\alpha)^{1/2}\Omega_\bot$.
Since the right-hand side of this inequality is smaller than $m\Omega-n^{1/2}\Omega_\bot$, we expect that
the eigen-frequency of the trapped oscillations become smaller than that in the case of no magnetic
fields, if other parameters are fixed (see Kato 2011a).
The frequency of the trapped oscillations depends also on the node number of oscillations in the radial direction
(say, $n_{\rm r}$), and decreases with increase of $n_{\rm r}$ if other parameters are fixed (see Kato 2011a).

In this context, we examine in this paper whether the following picture can describe the observed frequency-correlation 
of twin kHz QPOs with reasonable parameter values.
That is, we regard the fundamental mode of the vertical p-mode oscillations 
with no node ($n_{\rm r}=0$) in the radial direction as the upper kHz QPO, and the first overtone 
($n_{\rm r}=1$) in the radial direction as the lower kHz QPO.
Then, we examine whether the correlated time variation of the above two oscillations ($n_{\rm r}=0$ and $n_{\rm r}=1$)
due to time variation of magnetic fields can describe the correlated time variation of observed twin QPOs.
It is noted that time variation of global magnetic fields will be natural, since they are amplified by 
winding due to differential rotation and decreases by dissipation due to reconnection.

\section{Calculation of Frequencies of Trapped Vertical p-mode Oscillations}

Frequencies and their parameter dependences of trapped vertical p-mode oscillations in disks with toroidal magnetic 
fields have been numerically calculated by Kato (2011a) by a perturbation method. 
Since we use here the results, we briefly summarize the procedures adopted by Kato (2011a).

The Newtonian formulation is adopted except that the effects of general relativity are taken into account
when we condiser the radial distributions of $\Omega(r)$, $\kappa(r)$, and $\Omega_\bot(r)$.
The angular velocity of disk rotation, $\Omega(r)$, is approximated by the relativistic Keplerian one, since 
geometrically thin disks are considered.
We adopt the cylindrical coordinates ($r$, $\varphi$, $z$), where the $z$-axis is perpendicular to the 
unperturbed disk plane and the origin is at the disk center.

\subsection{Vertical Structure of Unperturbed Disks with Toroidal Magnetic Fields}

The unperturbed disks are assumed to be axisymmetric with toroidal magnetic fields.
The fileds are assumed to be purely toroidal with no poloidal component:
\begin{equation}
    \mbox{\boldmath $B$}_0(r,z)=[0,B_0(r,z),0].
\label{2.1}
\end{equation}
We further assume that the gas is isothermal in the vertical direction and $B_0(r,z)$ is distributed in such a way
that the Alfv\'{e}n speed, $c_{\rm A}$, is constant in the vertical direction, i.e., $(B_0^2/4\pi\rho_0)^{1/2}=$
const. in the vertical direction, where $\rho_0(r,z)$ is density in the unperturbed disks.

Then, the hydrostatic equilibrium in the vertical direction gives that $\rho_0(r,z)$ and $B_0(r,z)$ are 
distributed in the vertical direction as
\begin{equation}
    \rho_0(r,z)=\rho_{00}\ {\rm exp}\biggr(-\frac{z^2}{2H^2}\biggr)\quad {\rm and} \quad
    B_0(r,z)=B_{00}\ {\rm exp}\biggr(-\frac{z^2}{4H^2}\biggr),
\label{3.2}
\end{equation}
where the scale heigth, $H$, is related to $c_{\rm s}$, $c_{\rm A}$, and $\Omega_\bot$ by
\begin{equation}
     H^2(r)=\frac{c_{\rm s}^2+c_{\rm A}^2/2}{\Omega_\bot^2}.
\label{3.3}
\end{equation}

\subsection{Calculations of Frequencies of Trapped Oscillations}

Small-amplitude perturbations are superposed on the  equilibrium disks described above.
The perturbations are taken to be proportional to exp$[i(\omega t-m\varphi)]$.
Then, equations describing the perturbations are partial differential equations 
with respect to $r$ and $z$ [see equation (20) by Kato 2011a].
The equation is solved by being approximately decomposed into an equation describing the
behaviour in the vertical direction and that in the radial direction.
When doing so, we remember that we are now interested in the vertical p-mode oscillations, and thus
among three velocity components of $u_r$, $u_\varphi$, and $u_z$, associated with the perturbations 
the main one is $u_z$.
Furthermore, in the lowest order of approximations, the $z$-dependence of $u_z$ can be 
expressed in terms of the Hermite polynomials ${\cal H}_{n-1}(z/H)$, where $n$ is a positive integer
representing node number of oscillations in the vertical direction (Okazaki et al. 1987).
Considering these facts, and using a perturbation method, we derive an ordinary differential equation
with respect to $r$, which describes the behavior of perturbations in the radial direction.
The equation is solved by a WKB method with relevant boundary conditions in the radial direction to obtain
the eigen-frequency, $\omega$, of trapped oscillations.
As an inner boundary condition to solve the equation we adopt that at the inner edge of the disk
(which is taken at the radius of marginary stable circular orbit) $u_z$ vanishes, 
and as the outer boundary condition we assume that the outside of the trapped region (the evanescent region), 
$u_z$ decreases outwards.
The detailed procedures and the results on frequencies of trapped oscillations are given by Kato (2011a).

The frequencies of trapped oscillation depend on various parameters specifying the disk structure
(as well as the central sources) and the modes of oscillations.
As the former types of parameters we have $c_{\rm A}^2/c_{\rm s}^2$, $M$ (mass of the central source), and $a_*$ 
(spin parameter).
As the latter ones, we have $m$, $n$, and $n_{\rm r}$.
As mentioned before, we consider two-armed oscillations, i.e., $m=2$.
Concerning $n$ and $n_{\rm r}$, we restrict our attention in this paper to $n=1$ and 2 (i.e., the fundamental and first overtone
in the vertical direction), and $n_{\rm r}=0$ and 1 (i.e., the fundamental and first overtone in the radial direction).

\section{Frequency Changes and Their Correlation}

By using the procedures mentioned in the above section, Kato (2011a) numerically calculated frequencies of the trapped
vertical p-mode oscillations for various parameters.
We arrange the results from the viewpoint how the frequency change of the $n_{\rm r}=0$ oscillation and that of 
the $n_{\rm r}=1$ oscillation are correlated, when $c_{\rm A}^2/c_{\rm s}^2$ is changed, under 
other parameters being fixed.
The correlation is examined to two different oscillation modes, i.e., $n=1$ and $n=2$.
Figure 1 is for $n=1$, while figure 2 is for $n=2$.

The correlation depends on $M$ and $a_*$.
In figure 1, two cases of $M=1.4 M_\odot$ and  $M=1.8M_\odot$ are shown. 
The cases of $M=1.4M_\odot$ are shown by thick curves for two different values of spin parameter, i.e., $a_*=0$ and 
$a_*=0.2$.
The thin curves are for $M=1.8M_\odot$ and two cases of $a_*=0$ and $a_*=0.2$ are shown.
The dots attached on each curves show the value of $c_{\rm A}^2/c_{\rm s}^2$.
When $c_{\rm A}^2/c_{\rm s}^2$ is small, both frequencies of the $n_{\rm r}=0$ and $n_{\rm r}=1$ oscillations
are high and the point for the $c_{\rm A}^2/c_{\rm s}^2$ is on the upper-right corner of the correlation curve.
As $c_{\rm A}^2/c_{\rm s}^2$ increases, the point move towards lower-left corner 
along the curve, as shown in the figure.
It is noted that the correlation curves shift downward as $M$ increases, while upward as $a_*$ increases.

In figure 2, the frequency-frequency correlation is shown for oscillations with $n=2$, as mentioned before.
The set of ($M$, $a_*$) adopted is, however, different from that in figure 1.
That is, the thick curves are for $M=1.4M_\odot$ as in figure 1, but
the values of $a_*$ adopted are $a_*=0$, $0.4$, and $0.6$.
The thin curves are for $M=1.2M_\odot$, and the value of $a_*$ adopted are $a_*=0$ and 0.4. 

One of major differences between oscillations of $n=1$ and $n=2$ is that 
if the value of $c_{\rm A}^2/c_{\rm s}^2$ is the same, the oscillations of $n=2$ (figure 2) have lower
frequencies compared with those of $n=1$ (figure 1).
Dependences of correlation curves on $M$ and $a_*$ are qualitatively the same in both cases of $n=1$ (figure 1) and $n=2$
(figure 2).

%Figure 1
\begin{figure}
  \begin{center}
    \FigureFile(80mm,80mm){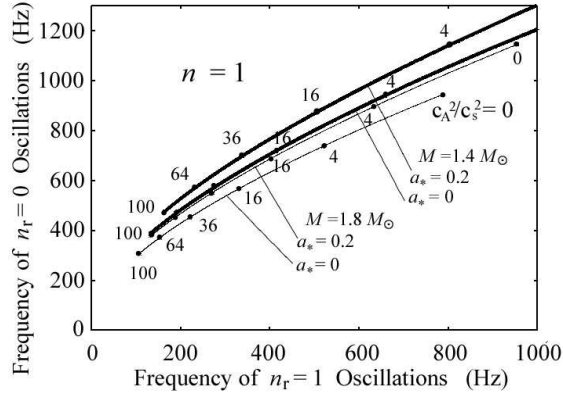}
    %%% \FigureFile(width,height){filename}
  \end{center}
  \caption{
The frequency correlation between the $n_{\rm r}=0$ and $n_{\rm r}=1$ oscillations, when $c_{\rm A}^2/c_{\rm s}^2$
is changed.
This figure is drawn for oscillations with $n=1$.
The correlation depends on $M$ and $a_*$.
The thick curves are for $M=1.4M_\odot$ and two cases of $a_*=0$ and $a_*=0.2$ are shown.
The thin curves are for $M=1.8M_\odot$ and two cases of $a_*=0$ and $a_*=0.2$ are shown.
The dots attached on curves are the value of $c_{\rm A}^2/c_{\rm s}^2$.
}
\label{fig:figure 1}
\end{figure}

%Figure 2
\begin{figure}
  \begin{center}
    \FigureFile(80mm,80mm){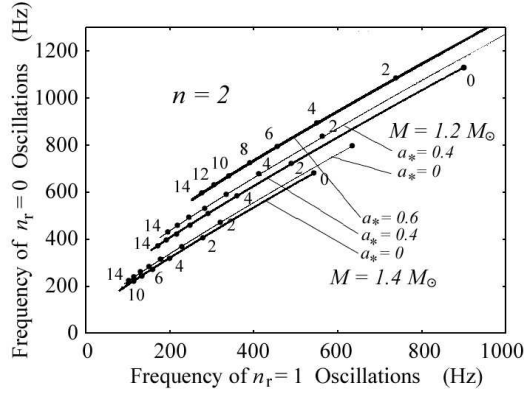}
    %%% \FigureFile(width,height){filename}
  \end{center}
  \caption{
The frequency correlation between the $n_{\rm r}=0$ and $n_{\rm r}=1$ oscillations, when $c_{\rm A}^2/c_{\rm s}^2$
is changed.
This figure is drawn for oscillations with $n=2$.
The thick curves are for $M=1.4M_\odot$ and three cases of $a_*=0$, $a_*=0.4$, and $a_*=0.6$ are shown.
The thin curves are for $M=1.2M_\odot$ and two cases of $a_*=0$ and $a_*=0.4$ are shown.
The dots attached on curves are the values of $c_{\rm A}^2/c_{\rm s}^2$ as in figure 1.
}
\label{fig:figure 2}
\end{figure}

\section{Comparison with Observed Frequency Correlation}

The next problem is to compare the frequency-frequency correlation shown in figures 1 and 2
with the observed correlation between the lower and upper kHz QPOs.
The latter has been plotted on a frequency-frequency diagram
for some typical neutron-star X-ray sources (e.g., Abramowicz 2005; Belloni et al. 2007).
Here we adopt the figure by Abramowicz (2005) and superpose it on figures 1 and 2, which are shown respectively as figures 3
and 4.
In order to avoid complexity, the values of $c_{\rm A}^2/c_{\rm s}^2$ as well as $M$ and $a_*$ are not shown 
in the superposed figures.

Comparison of figures 3 and 4 shows that the correlation 
in the case of $n=1$ (figure 3) better fits to the observed correlation of twin kHz QPOs,  
compared with that of $n=2$ (figure 4).
For example, in the case of $n=1$, the observed correlation points of Sco X-1 (green points) are well 
on the correlation curve calculated by $M=1.4M_\odot$ and $a_*=0$ throughout the whole observational points.
In this case of $M$ and $a_*$, the range of variation of $c_{\rm A}^2/c_{\rm s}^2$ necessary to describe the 
observations of Sco X-1 is roughly $c_{\rm A}^2/c_{\rm s}^2=2\sim 6$.
The set of ($M$, $a_*$) that well describe observations are, however, not unique.
In the case of Sco X-1, for example, we can describe observational data by adopting a higher mass than 
$1.4M_\odot$ and a non-zero $a_*$.
That is, if we adopt $M=1.8M_\odot$, for example, the curve of $a_*\sim 0.25$ fits well the observational data (figure 3).

One of reasons why the $n=1$ oscillations will be more realistic than the $n=2$ ones is that in the latter case 
a rather high spin is required to describe observations if we take $M>1.4M_\odot$ (figure 4).
In other words, $M<1.4M_\odot$ is required if the spin is taken to be moderately small, say $a_*<0.2$ (see figure 4).

If we take the picture that the observed twin kHz QPOs are the $n=1$ oscillations, a natural question raised 
is whether the $n=2$ oscillations are observed or not.
One of possibilities is that the $n=2$ oscillations are not observed since their amplitude will be small.
The $n=2$ oscillations have one more node in the vertical direction compared with the $n=1$ oscillations.
Hence, the saturated amplitude of the $n=2$ oscillations
may be smaller than that of the $n=1$ oscillations.\footnote{
The saturated amplitude will be determined by the balance between viscous damping of oscillations 
and stochastic excitation by turbulence.
The viscous damping will become strong as $n$ increases, since the vertical wavelength becomes short with increase of $n$.
} 
Another possibility is that the QPOs observed in a low frequency region (say, 150 Hz $\sim$ 300 Hz) are the $n=2$ oscillations.
This is because if we want to describe low frequency QPOs by the $n=1$ oscillations, too strong magnetic fields
are required (see figures 1 and 3).

%Figure 3
\begin{figure}
  \begin{center}
    \FigureFile(120mm,120mm){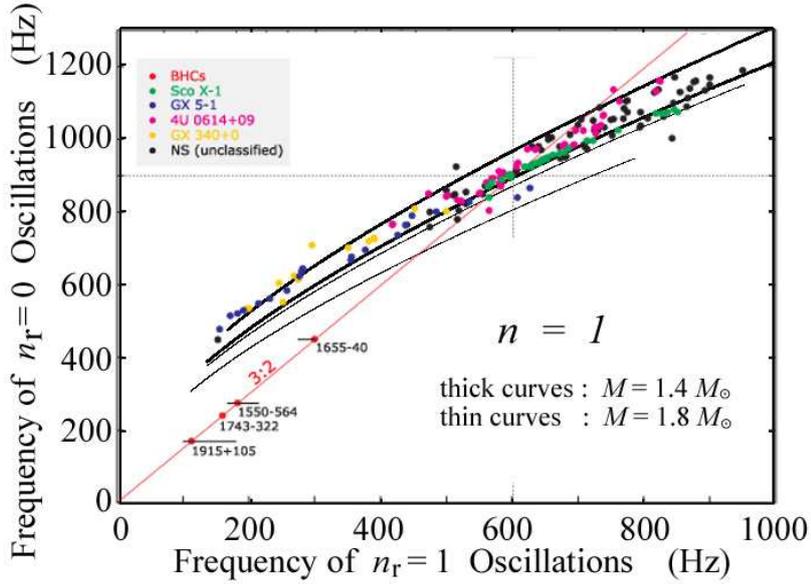}
    %%% \FigureFile(width,height){filename}
  \end{center}
  \caption{
Diagram comparing the calculated frequency-frequency correlation with the observed 
frequency correlations in some typical neutron-star X-ray sources.
The oscillations with $n=1$ are adopted.
This figure is a superposition of figure 1 and the observational data plotted by Abramowicz (2005).
The straight line labelled by 3:2 is the line on which the frequency ratio of twin 
QPOs is 3:2.
The four sources on the line are black-hole candidates and are outside of
our present concern.
}
\label{fig:figure 3}
\end{figure}

%Figure 4
\begin{figure}
  \begin{center}
    \FigureFile(120mm,120mm){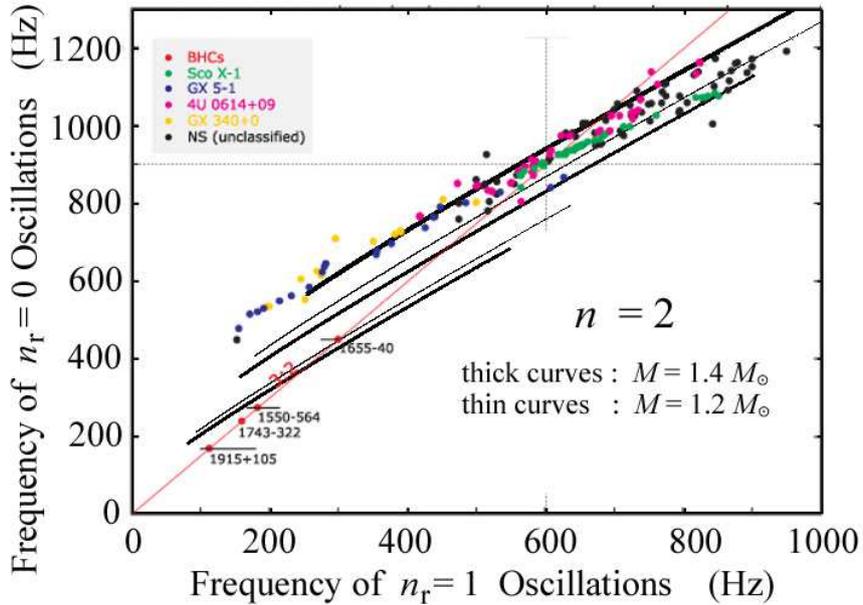}
    %%% \FigureFile(width,height){filename}
  \end{center}
  \caption{
The same as figure 3 except that $n=2$ oscillations are considered here.
That is, this is a superposition of figure 2 and the observational data plotted by Abramowicz (2005).
}
\label{fig:figure 4}
\end{figure}

\section{Summary and Discussion}

In this paper we take the standpoint that the kHz QPOs observed in neutron-star X-ray binaries are disk
oscillations trapped in the innermost region of relativistic disks.
There are many kinds of trapped oscillations in the inner region of disks, if the disks have inner boundary.
One of important checks examining which trapped oscillations can become a candidate of QPOs is
to study whether they can describe the frequency correlation of observed twin QPOs.
This problem, however, seems not to be examined quantitatively yet.
In this paper, we examined this problem to the vertical p-mode oscillations.
The results show that in the conventional values of parameters of mass and spin, 
the vertical p-mode oscillations seem to well describe the observed correlation of kHz QPOs.
%This suggests that more quantitative examination of frequencies of trapped vertical p-mode oscillations
%is worthwhile.

The magnetic fields required in this model, however, are rather strong, i.e., $\beta$ defined by 
$\beta\equiv p_{\rm g}/p_{\rm mag}=2c_{\rm s}^2/c_{\rm A}^2$ is less than unity, where $p_{\rm g}$ and $p_{\rm mag}$ 
are, respectively, the gas and magnetic pressures.
It should be noted that the presence of steady low-$\beta$ (i.e., $\beta<1$ or $c_{\rm A}^2/c_{\rm s}^2>2$)
disks has been numerically demonstrated by Machida et al. (2006) and Oda et al. (2007, 2009, 2010).
One of reasons why magnetically supported disks are possible is that the escape of magnetic fields from the 
disks by the Parker instability is suppressed by strong magnetic tension (Shibata et al. 1990).
A low-$\beta$ disk is a bridge between ADAF and slim disk, describing a bright hard state (optically thin)
and  a high/soft state (optically thick) (Oda et al. 2009, 2010).
In their numerical models, such a high value of $c_{\rm A}^2/c_{\rm s}^2$ as $c_{\rm A}^2/c_{\rm s}^2=100$
has been realized in extreme cases.
As far as the geometrically thin approximation is concerned, such a high value of $c_{\rm A}^2/c_{\rm s}^2$
is acceptable.
For example, for $c_{\rm A}^2/c_{\rm s}^2=50$, equation (\ref{3.3}) gives $H/r\sim 5c_{\rm s}/r\Omega$.
In the standard disks, the value of $c_{\rm s}/r\Omega$ is about $1/30$ in the inner region of disks
(e.g., see figure 3.4 by Kato et al. 2008).
In low-$\beta$ disks, however, the temperature is higher than that of the standard disks by the presence 
of dissipation of strong magnetic fields, and is between the temperatures of ADAFs and of the standard disks.
If we take a temperature 10 times higher than that of the standard disks, we have $c_{\rm s}/r\Omega\sim 1/10$
and $H/r\sim 0.5$ for $c_{\rm A}^2/c_{\rm s}^2=50$.

In this paper, we assume that the disk is isothermal in the vertical direction.
If the vertical structure of the disks is not isothermal but polytropic, the polytropic index is an additional
parameter contributing to frequencies of trapped oscillations (Kato 2010).
That is, the frequencies of trapped oscillations depend not only on the strength of toroidal fields, but
also on the polytropic index.
The frequencies of trapped oscillations also depend on boundary conditions (see Kato 2011a).
This suggests the possibility that we can get more detailed information of the innermost region of disks by
detailed adjustment of calculated correlation curves to observational ones for individual sources,
if the present model of QPOs is correct.

It should be noted here that the trapped region of the vertical p-mode oscillations is rather narrow in
the case of $n=1$ and $n=2$, i.e., the outer edge of the trapped region is around 
$4r_{\rm g}$ when $a_*=0$ (see figure 1 by Kato 2011a).
A problem is whether the oscillations in such a narrow region can produce observed luminosity variations.
One of another problems related to the present model is that the magnetic fields required to describe the
low frequency QPOs (100 Hz $\sim$ 300Hz) is rather high (e.g., $c_{\rm A}^2/c_{\rm s}^2\sim 10^2$),
if the QPOs are regarded as the $n=1$ oscillations.
This may suggest that the sources with such low frequency QPOs have really strong magnetic fields or these QPOs 
are oscillations of $n=2$.
%To do more careful comparison with models and observed oscillations, it will be worthwhile comparing the 
%calculated curves with the observational data of individual sources 
%rather than those of assembly of many sources.

Whether trapped vertical p-mode oscillations are really excited in disks is a problem remained to be examined.
We suppose that they will be excited by stochastic processes of turbulence in disks, as solar and stellar 
non-radial oscillations are now known to be excited by the processes in the convection zone.
Turbulence has two effects on oscillations. 
One is their excitation by stochastic processes and the other is their damping by usual turbulent viscous processes.
%In the case of disks, the excitation process will be more effective than in stars, since the turbulence (MHD turbulence)
%in disks is much stronger than in stars.
In the case of disks, the turbulence is MHD turbulence resulting from the magneto-rotational instability (MRI),
and is much stronger than that in stellar convection zone.
Hence, the turbulence  will excit and maintain trapped oscillations at a high level against dissipation.

%Observations show that the frequency correlation of the twin kHz QPOs is phenomenologically extended to dwarf novae 
%(Warner et al. 2003).
%In our present model of correlation of kHz QPOs, the general relativity is not an important ingredient, except that
%the Keplerian rotation is relativistic.
%This suggests the possibility that the present model can be qualitatively extended to QPOs in dwarf novae disks.

Finally, high-frequency twin QPOs observed in black-hole X-ray sources are briefly mentioned.
These QPOs have robust frequencies with their frequency ratio 3:2.
Hence, the present model cannot describe these QPOs, unless in the black-hole cases we have particular processes 
which keep the strength of magnetic fields unchanged with time by balance among creation 
(by differential rotation and MRI), escape (by jet and accretion), and dissipation (reconnection) of magnetic fields.
In addition, the frequency ratio must be kept to 3:2.
Considering these situations, we think that some additional mechanisms will be necessary to describe the twin QPOs 
of black-hole X-ray binaries in the framework of the present model of trapped oscillations.

\bigskip
\leftskip=20pt
\parindent=-20pt
\par
{\bf References}
\par
%Abramowicz, M. A., \& Klu{\' z}niak, W. 2001, A\&A, 374, L19 \par
Abramowicz, M.A. 2005, AN, 326,782 \par
Belloni, T., M{\' e}ndez, \& Homan, J. 2007, MNRAS, 376,1133 \par
Ferreira, B.T. \& Ogilvie, G.I. 2008, MNRAS, 386, 2297 \par
Fu, W. \& Lai, D. 2009, ApJ, 690, 1386\par
Kato, S. 1990, PASJ, 42, 99 \par
Kato, S. 2001, PASJ, 53, 1\par 
%Kato, S. 2003, PASJ, 55, 257 \par
Kato, S. 2004, PASJ, 56, 905 \par
Kato, S. 2007, PASJ, 59, 451 \par
Kato, S. 2008a, PASJ, 60, 111\par
Kato, S. 2008b, PASJ, 60, 1387\par
Kato, S. 2010, PASJ, 62, 635\par
Kato, S. 2011a, PASJ, 63, 125, \par
Kato, S. 2011b, PASJ, 63, in press, arXiv:1103.0677 [astro-ph.HE] \par
%Kato, S., Fukue, J., \& Mineshige, S. 1998, Black-Hole Accretion Disks 
%  (Kyoto: Kyoto University Press)\par
Kato, S. \& Fukue, J. 1980, PASJ, 32, 377\par
Kato, S. \& Fukue, J. 2006, PASJ, 58, 909 \par  
Kato, S., Fukue, J., \& Mineshige, S. 2008, Black-Hole Accretion Disks 
  -- Toward a New paradigm -- (Kyoto: Kyoto University Press)\par
Kato, S., Okazaki, A.-T., Oktariani, F. 2011, PASJ, 63, in press, arXiv:1101.1593 [astro-ph.HE]\par
%Kato, Y., Miyoshi, M., Takahashi, R., Negoro, H., \& Matsumoto, R. 2010, MNRAS, 403L, 74\par
%Klu{\' z}niak, W., \& Abramowicz, M. 2001, Acta Phys. Pol. B32, 3605   \par
Lai, D., \& Tsang, D., 2009, MNRAS, 393, 979\par
%Li, L.-X., Goodman, J., Narayan, R. 2003, ApJ, 593,980 \par
Lin, Y.-F., Boutelier, M., Barret, D., Zhang, S.-N. 2010, ApJ, 726,74\par
Machida, M., Nakamura, K.E., \& Matsumoto, R. 2006, PASJ 56, 671\par
Nowak, M.A., \& Wagoner, R.V. 1992, ApJ, 393, 697\par
%McClintock, J.E., Remillard, R.A. 2005, "Black Hole Binaries", in
%   Compact Stellar X-ray Sources, eds. W.H.G. Lewin and M. van der Klis,
%   Cambridge University Press, Cambridge, in press; astro-ph/0306213 \par
%McClintock, J.E., Shafee, R., Narayan, R., Remillard, R.A., Davis, S.W., \& Li,L. 2006, ApJ, 652,518\par
%Middleton, M., Done, C., Gierli\'{n}ski, M., Davis, S.W. 2006, MNRAS, 373, 1004\par
Oda, H., Machida, M., Nakamura, K.E., \& Matsumoto, R. 2007, PASJ, 59, 457\par
Oda, H., Machida, M., Nakamura, K.E., \& Matsumoto, R. 2009, ApJ, 697, 160\par
Oda, H., Machida, M., Nakamura, K.E., \& Matsumoto, R. 2010, ApJ, 712, 6390\par
Okazaki, A.-T., Kato, S., \& Fukue, J. 1987, PASJ, 39, 457\par
Oktarian, F., Okazaki, A.-T., \& Kato, S. 2010, PASJ, 62,709\par
Ortega-Rodr\'{i}gues, M., Silbergleit, A.S., \& Wagoner, R.V. 2002, ApJ. 567, 1043\par
Perez, C.A., Silbergleit, A.S., Wagoner, R.V., \& Lehr, D.E., 1997, ApJ, 476, 589\par
%Remillard, R.A. 2005, Astron. Nachr. 326, 804 \par
%Shafee, R., McClintock, J.E., Narayan, R., Davis, S.,W., Li, L., \& Remillard, R.A. 2006, ApJ., 636, L113\par
Shibata, K., Tajima, T., \& Matsumoto, R. 1990, ApJ, 350, 295\par
Silbergleit, A.S., Wagoner, R.V., \& Ortega-Rodr\'{i}gues, M. 2001, ApJ., 548, 335\par
%Stella, L., and Vietri, M. 1999, Phys. Rev. L82, 17 \par 
%Steiner, J.E.,Reis, R.C., McClintock, J.E., Narayan, R., Remillard, R.A., Orosz, J.A., Gou, L., Fabian, A.C., \& Torres, M.A.P.
%    2010, MNRAS, 2010, arXiv:1010.1013v2 [astro-ph.HE]\par
%Tsang, D., \& Lai, D. 2009a, MNRAS, 393, 992\par
%Tsang, D., \& Lai, D. 2009b, MNRAS, 400,470\par
%T{\"o}r{\"o}k, G., 2007, private communication \par
%van der Klis, M. 2004,     \par
Wagoner, R.V. 1999, Phys. Rep., 311, 259\par
%Warner, B., Woudt, P.A., \& Pretorius, M.L. 2003, MNRAS, 344,1193\par
%Wijnandns, R., van der Klis, M., Homan, J., Chakrabarty, D., Markwardt, C.B., \& Morgan, E.H. 2003, Nature, 424, 44 \par
%Zhang, C.M., Wang, J., Zhao, Y.H., Yin, H.X., Song, L.M., Menezes, D.P., 
%   Wickramasinghe, D.T., Ferrario, L., \& Chardonnet, P. 2010,
%   astro-ph.HE, arXiv:1010.5429v4  \par 

\end{document}